# Metastability from Photoluminescence of *n*-type GaN

C. S. Park and T. W. Kang

*Research Institute for Natural Science and Quantum-functional Semiconductor Research Center,*

*Dongguk University, 3-26, Pil-dong, Chung-ku, Seoul, Korea*

We measured the temperature dependence of photoluminescence involved with the metastability of unintentionally doped GaN. Reaction energy *U* of donor atom is 0.1eV and shallow donor is more stable than deep center. The impurity transition was applied to unintentionally doped GaN at low temperature and reaction energy *U* was provided for shallow-deep transition. We propose that the origin of *DX* center in unintentionally doped wurtize GaN is considered to be an oxygen impurity instead of silicon.

## INTRODUCTION

GaN has been issued as an ideal material for the optical and electrical devices because of direct wide band gap semiconductor[1,2] and thermal stability gives rise to attention on high temperature and high power devices. As-grown GaN has usually *n*-type conductivity with high background concentration, $\sim 10^{19}/cm^3$. Native defect and impurities during the growth were proposed for the candidates of the high electron density in unintentionally doped GaN. Especially, *DX* center formation has been reported in the case of impurities of donor[3-6]. Deep donor levels have been studied for several decades of years because of their unusual characteristics and strange effects on the properties of III-V semiconductors. McCluskey et al[7], discovered a metastability in $Al_xGa_{1-x}N$ that can be ascribed to oxygen donors. The increase in the donor binding energy with x is consistent with a localized *DX* state which interests the conduction band at x=0.27. In $Al_{0.44}Ga_{0.56}N$ intentionally doped with Si, no evidence was observed for donor metastability, in agreement with theoretical calculations which predict that silicon does not form *DX* centers in AlGaN[7]. First principle calculations[8] reported that oxygen can occupy a substitutional nitrogen site($O_N$) and act as a shallow donor, with a low formation energy under typical growth conditions. In $Al_xGa_{1-x}As$ alloys with x>0.22, the *DX* center is the lowest-energy state of silicon. Chadi and Chang proposed that *DX* center is a negatively charged and highly localized defect center, in which the Si atom is displaced into an interstitial position, resulting from a large lattice distortion[9].

High concentrations of *DX* levels occur in many device structures. It is now generally believed that the *DX* level is a state of the isolated substitutional donor. The variation of the transport properties and capture and emission kinetics of the **DX** level with the conduction-band structure is now well understood. Variations in the local environment of the donor atom have been shown to perturb the thermal emission kinetics. It has been known that the lattice relaxation which occurs when electrons are trapped involves the motion of an atom from a group-III site to an interstitial site[10, 11].

In this paper, the measurement of photoluminescence (PL) was performed on the unintentionally doped GaN grown by Plasma Assisted Molecular Beam Epitaxy (PAMBE) in order to investigate the optical properties. We report newly that at low temperature, unidentified anomalous peaks appear in the results of temperature-dependent photoluminescence for unintentionally doped GaN and the reaction energy **U** of a donor atom is proposed for the explanation of shallow-deep transition.

EXPERIMENTAL DETAILS

We have grown GaN epilayer on $Al_2O_3(0001)$ substrate by PAMBE with two step buffer layer. An inductively coupled radio frequency nitrogen plasma source with a purity of 99.9999% provided the reactive nitrogen, and the Ga were evaporated using conventional effusion cells. As soon as the chemical cleaning process was finished, the sapphire substrate was mounted onto a molybdenum susceptor. Prior to GaN growth, the substrate surface was exposed to an activated nitrogen beam for 10 min for complete covering with a nitridated layer. The deposition of the GaN active epilayer on the GaN buffer layer grown at 550℃ was done at a substrate temperature 750℃. The thickness of the GaN epilayer was 750nm. During the growth, the crystalline quality of the GaN epitaxial layers was monitored in situ by reflection high energy electron diffraction (RHEED). Van der Pauw Hall effect measurements were carried out at room temperature in a magnetic field of 0.5T by using a Keithley 181 nanovoltmeter. The Photoluminescence(PL) measurements were performed using a 75 cm monochromator equipped with a GaAs photomultiplier tube. The excitation source was the 3250Å line of a He-Cd laser, and the sample temperature was controlled between 10 and 240K by using a He displex system.

RESULTS AND DISCUSSION

The grown sample on sapphire substrate is an unintentionally doped *n*-type GaN. Hall effect measurements show that the carrier concentration is $1.6 \times 10^{17}/cm^3$. Fig. 1 shows the results of temperature-dependent photoluminescence and reveals the peculiar properties of unidentified peaks around 3.32eV according to the temperature variation. The intensity of free exciton peak is higher than the bound exciton peak. Blue shifts of bound exciton peak take place below 70K and the peak position shifts to low energy region due to energy loss by phonon during the large lattice relaxation[11,12]. The difference between neutral donor bound exciton and ionized donor bound exciton is 11.5 meV which is same in this measurement[13]. Fig. 2 shows that one of the important properties happen in the temperature dependence of photoluminescence measurement. Shallow and deep defect levels coexist at low temperature and interacts each other. The only shallow donor level exists at 180K. As the temperature increases over 150K, the shallow donor moves into the deep position at the range 180K to 210K. Then shallow donor level disappears. In our result, it is explained that there is a tunneling mechanism of free carriers through radiative recombination via deep levels for the reason of low intensity. Carriers may move to another radiative level($A^*$) instead of recombination in shallow level($A$) and possible vice versa. The intensity of transition level is related to the decreases of dopant concentration[14]. The distribution of the electrons occupying the $A$ and the $A^*$ states of oxygen is dependent on the temperature. When the capture cross section of level $A$ is large enough to catch all the electron that tunnel from level $A^*$, the increase in the number of neutral donors is determined from the number of the electrons that tunnel to level $A$. The number of electrons transferred from $A$ to $A^*$ can be ignored. The variance of the number of electrons occupying the level $A$ ( $\Delta n$ ) as a function of temperature ($T$) is given by

$$\begin{aligned}
\Delta n &= N_{A^*}C_n[1 - \exp(-E_{A^*}/k_BT)] \\
&- N_{A^*}P_T[1 - \exp(-E_{A^*}/k_BT)],
\end{aligned} \tag{1}$$

where $N_{A^*}$ is the carrier concentration of the trap level $A^*$, $C_n$ is the capture coefficient of the trap level $A^*$, $k_B$ is the Boltzman's constant, and $P_T$ is the tunneling probability. The tunneling probabilities increase with increasing temperature. While the dominant process at low temperatures is electron trapping, that at relatively high temperatures is the transport of electron hopping through the tunneling process.

This result is similar to the **DX** center in GaAs, Al$_x$Ga$_{1-x}$As alloy. Shallow-deep transition in temperature-dependent PL spectra means the existence of defect center interacting with shallow donor level. Reaction process of a donor in semiconductor explains that a donor atom is ionized with the large lattice relaxation [10]. The reaction process is applied to unintentionally doped GaN and positive **U** energy is supplied for shallow-deep transition[9]. The reaction mechanism is as follows

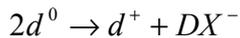

$$2d^0 \rightarrow d^+ + DX^-$$

the capture barrier of **DX** formation is obtained to be 0.51eV from the activation energy calculation of the temperature-dependent photoluminescence of Fig. 3. Furthermore, the result of Fig. 1 demonstrates the positive **U** energy is 0.1eV from the difference between shallow levels and deep levels. The experimental result is well consistent with the report of AlGaN:O by Mckluskey *et al*[7]. Theoretical calculation for negative **U** energy in WZ-GaN is -0.48eV and the state is unstable, which was done by first principle method[8]. The shallow-deep transition can be explained from Fig. 3. The equation of transition is as follows

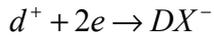

This implies that impurities were doped unintentionally during the growth. It is considered that the origin of *n*-type conductivity in unintentionally doped GaN is not native defects but impurities during the growth.

## SUMMARY AND CONCLUSIONS

In summary, we measured the temperature-dependent photoluminescence of GaN grown by PAMBE. The metastability between shallow donor and deep center exists in unintentionally doped GaN grown by PAMBE. These phenomena can be explained by reaction energy $U$ as in GaAs and $Al_xGa_{1-x}As$. Positive $U$ energy is 0.1eV. The capture barrier of DX center is 0.51eV. Theoretical calculation for the *DX* center formation of oxygen donor in GaN is consistent with our $U$ energy. It is suggested that the origin of unintentionally doped high background concentration is attributed to oxygen atom instead of silicon.

## ACKNOWLEGEMENT

This work was supported by the Korea Science and Engineering Foundation through the Quantum-Functional Semiconductor Research Center(QSRC) and by a research fund of the Dongguk University.

Figure Caption

Fig.1 Photoluminescence spectra measured at several temperatures for unintentionally doped GaN epilayers grown on sapphire substrates.

Fig.2 Photoluminescence spectra in the energy range between 3.1 and 3.6eV measured at several temperatures for unintentionally doped GaN epilayers grown on sapphire substrates. The measurement temperature is between 180 and 220K.

Fig.3 The energy of the donor bound exciton peak as a function of 1000/T, as obtained from Fig. 2

**Fig. 1**

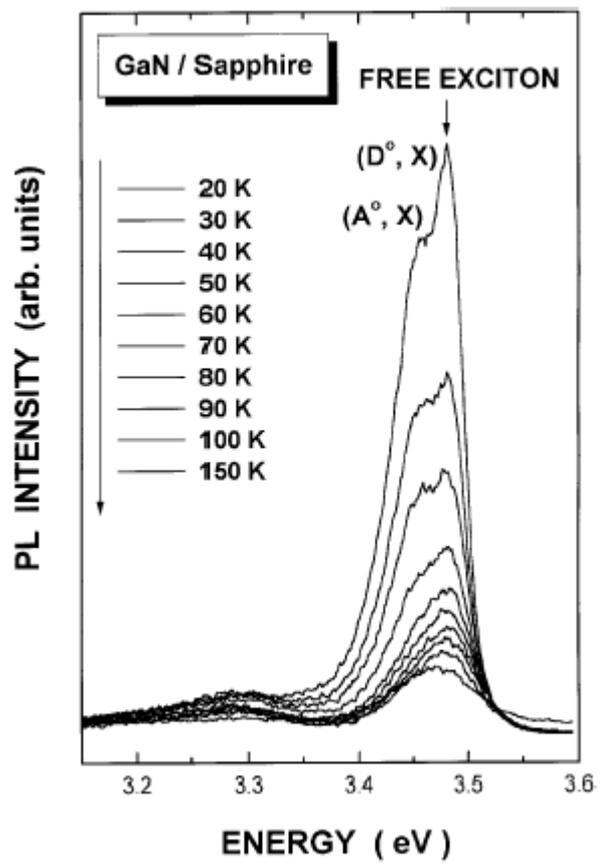

**Fig. 2**

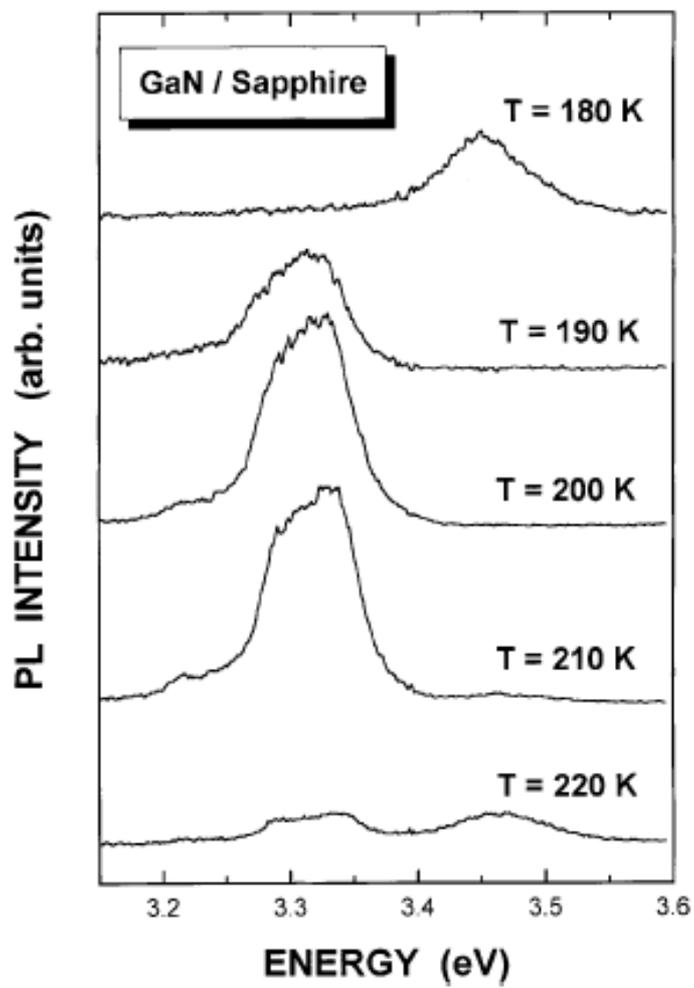

**Fig. 3**

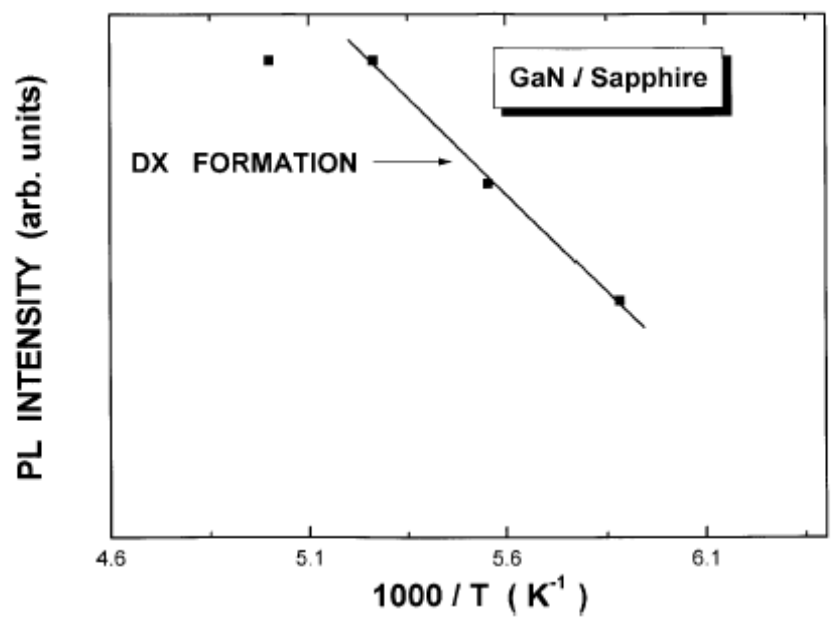